\documentclass[twocolumn,showpacs,preprintnumbers,amsmath,amssymb,prb,aps]{revtex4-1}
\usepackage{epsfig}
\usepackage{epstopdf}
\usepackage{graphicx}
\usepackage{dcolumn}
\usepackage{bm}
\usepackage{slashbox}
\usepackage{textcomp}
\usepackage{array}
\usepackage{subcaption}
\usepackage{booktabs}

\begin{document}

\title{A comparative study of different exchange-correlation functionals in understanding structural, electronic and thermoelectric properties of Fe$_{2}$VAl and Fe$_{2}$TiSn compounds}
\author{Shivprasad S. Shastri}
\altaffiliation{Electronic mail: shastri1992@gmail.com}
\author{Sudhir K. Pandey}
\affiliation{School of Engineering, Indian Institute of
Technology Mandi, Kamand - 175005, India}

\date{\today}

\begin{abstract}
Fe$_{2}$VAl and Fe$_{2}$TiSn are full Heusler compounds with non-magnetic ground state. The two compouds are good thermoelectric materials. PBE and LDA(PW92) are the two most commonly used density functionals to study the Heusler compounds. Along with these two well studied exchange-correlation functionals, recently developed PBEsol, mBJ and SCAN functionals are employed to study the two compounds. Using the five functionals equilibrium lattice parameter and bulk modulus are calculated. Obtained values are compared with experimental reports wherever available. Electronic structure properties are studied by calculating dispersion curves, total and partial density of states. For Fe$_{2}$VAl, band gap of 0.22 eV is obtained from the mBJ potential which is in reasonable agreement with experimental value while, for Fe$_{2}$TiSn band gap of 0.68 eV is obtained. Fe$_{2}$VAl is predicted to be semimetallic with different values of negative gaps from LDA,PBEsol,PBE and SCAN functionals. Whereas, Fe$_{2}$TiSn is found to be semimetallic(semiconducting) from LDA,PBEsol(PBE,SCAN) functionals employed calculations.   From the dispersion curve effective mass values are also computed to see the contribution to the Seebeck coefficient. In Fe$_{2}$TiSn, a flat band is present along the $\Gamma$-X direction  with calculated value of effective mass $\sim$36 more than the mass of electron. The improvements or inadequacies among the functionals in explaining the properties of full Heusler alloys for thermoelectric application are thus observed through this study. 

\end{abstract}

\maketitle

\section{Introduction} 
Electronic band structure of a material tells about the occupation of electrons at different energy levels in that material. The electronic structure can be studied through experimental methods like photoemission spectroscopy as well as through theoretical methods. Electrical conductivity, thermal conductivity, Seebeck coefficient are transport properties of a material. For a thermoelectric material, it's efficiency is determined by figure-of-merit(ZT) and it is governed by these transport properties.\cite{snyder1,snyder2} The transport properties can be well explained by understanding the electronic structure of the material. To improve and modify a material as an efficient thermoelectric of practical application a good understanding of it's electronic structure is essential. So, the methods of electronic structure analysis should be accurate enough to model the  material with physical accuracy.

	First-priniciples density functional theory(DFT)\cite{dfthk} method is the most resorted methods for the theoretical evalulation of electronic structure of periodic solids. In the Kohn-Sham(KS)\cite{kohnsham} form of DFT, the KS equation is solved self-consistently for the one electron wave functions. In the KS equation the electron-electron interaction part is approximated by the exchange-correlation potential. The limitation of DFT in exactly modelling electronic structure is introduced from this part. So, there is a large research in the development of electron exchange-correlation functionals for the better approximations. Thus, many density functionals exists today with it's own merits and demerits. The limitations of DFT functionals is that they may be quite accurate for some physical properties, while it may not be accurate for other physical properties of the same material.\cite{sholldft} Some functionals are constructed for explaining specific applications.

Local densiy approximation of Perdew and Wang-1992(LDA-PW92)\cite{lda92} and generalized gradient approximation of Perdew-Burke-Ernzerhof(GGA-PBE)\cite{pbe} are the two most widely used functionals in the first-principles DFT calculations. In the LDA, local exchange-correlation potential is defined as the exchange potential for the spatially uniform electron gas with the same density as the local electron density.\cite{sholldft} LDA is less suitable for prediction of the  properties of atoms and molecules, since the density is not slowly varying in case of atoms and molecules. In GGA functionals electron density is descibed using both the local electron density and gradient of electron density. PBE is an improved description of the local spin density approximation for atoms and molecules.\cite{pbe} GGA calculations are found to improve upon LDA for atomization of energies of molecules and enthalpy of formation derived from the atomization energy.\cite{bulkggalda09} But, for nonmolecular solids, the lattice parameters calculated by PBE not found to improve. Also, it is known that LDA underestimates lattice constant, while PBE overestimates. PBEsol functional was proposed by a restoration of the density-gradient expansion in PBE.\cite{pbesol} This functional is intedended to provide accurate values of equilibrium properties for solids and their surfaces. Also, PBEsol is supposed to give better values of lattice constant in the densly packed solids and in solids under pressure. mBJ potential was proposed by Tran and Blaha, by modifying the exchange potential originally put forth by Becke and Johnson. This semilocal potential is claimed to yield accurate band gaps for semiconductors and insulators with accuracy and it is less expensive than the hybrid and GW calculations.\cite{mbj} SCAN is a semilocal meta-GGA approximation which is fully constrained.\cite{scan} This functional is expected to be significant improvement over PBE, PBEsol and LDA functionals at the nearly same computational cost.

Number of studies on full Heusler alloys using LDA, PBE or mBJ functionals are available in literatures. But, a comparative study on full Heusler alloys, with different exchange-correlation functionals which influences in deciding the thermoelectric behavior is missing.   
Fe$_{2}$VAl and Fe$_{2}$TiSn are compounds belonging to class of full Heusler alloys with formula unit of the form $X_{2}YZ$, where $X$ and $Y$ are transition metal elements and $Z$ is a main group element.\cite{minert} The two compounds have non-magnetic ground state with Slater-Pauling rule for these full Heusler alloys giving zero magnetic moment.\cite{yabu} The two compounds are good thermoelectric materials. Their properties are studied experimentally as well as using first-principles calculations. Experimentally, Nishino \textit{et. al} reported that Fe$_{2}$VAl-based full Heusler alloys showing large power factor(PF=$S^{2}\sigma$, where $S$ is Seebeck coeffient and $\sigma$ is electrical conductivity) considerably more than that of the conventional thermoelectric material Be$_{2}$Ti$_{3}$.\cite{nishino06} In first-principles DFT based study of full Heusler compounds, LDA of Perdew Wang(1992), PBE are more commonly used functionals to investigate the thermoelectric properties. Markus Meinert investigated the properties of full and half Heusler alloys using modified Becke-Johnson potential .\cite{minert} Sharma \textit{et. al} used PBEsol exchange-correlation functional to study thermoelectric properties of the full Heusler alloys and showed the possibility of synthesis in laboratory.\cite{ssharma}

In the present work, taking up Fe$_{2}$VAl and Fe$_{2}$TiSn as representatives of non-magnetic class of full Heusler alloys we are interested to check the suitability of these exchange-correlation functionals for the calculation of the properties of full Heusler alloys with non-magnetic ground states. The properties approximated from the new functionals are compared with the well used functionals(LDA, PBE) used to study this kind of compounds. We have employed five exchange-correlation functionals  to study: i) structural properties of the two Heusler compounds. Lattice constant and bulk modulus values are extracted from energy versus volume curves and the obtained values are compared with the available experimental data.  ii) Dispersion curves, total and partial density of states are calculated to study the electronic structure. General features and differences in the electronic structure predicted from different functionals are discussed. For the two compounds effective mass values are computed from the dispersion curve using parabolic approximation. The values of effective mass are used to give an idea of contribution to Seebeck coefficient.

\section{Computational details}
The calculations are performed using the full-potential linearized augmented plane wave(FPLAPW) method as implemented in the WIEN2k\cite{wien2k} program for calculating crystal properties within density functional theory. For the exchange-correlation part five different functionals are used viz., LDA of Perdew-Wang-1992 (LDA)\cite{lda92}, GGA of Perdew-Burke-Ernzerhof(PBE)\cite{pbe}, and newly developed PBEsol\cite{pbesol}, mBJ\cite{mbj}, and SCAN\cite{scan}. In case of mBJ, for the correlation part LDA is  used with the modified Becke-Johnson(mBJ) potential for the exchange part. The muffin-tin radii $R_{MT}$ used for volume optimization caluclations of (i) Fe$_{2}$VAl are 2.26 bohr for Fe; 2.15 bohr for V and 2.04 bohr for Al (ii) Fe$_{2}$TiSn are 2.32 bohr for Fe;2.26 bohr for Ti; 2.32 bohr for Sn,  respectively. A k-mesh grid of size 10x10x10 is used for both volume optimization and electronic structure calcualtions. The self-consistency in the total energy/cell is achieved by setting a convergence criteria of 0.1 mRy.

The equilibrium lattice constants are computed by fitting the total energy versus volume of the unit cell data to the Birch-Murnaghan(BM) equation of state.\cite{bmeos} The third-order BM isothermal equation of state is given by the formula: 

\begin{align*}
  E(V)&=E_{0}+\frac{9V_{0}B_{0}}{16}\bigg[\big\{(\frac{V_{0}}{V})^{\frac{2}{3}}-1\big\}^{3}B_{0}' \\
  &+\big\{(\frac{V_{0}}{V})^{\frac{2}{3}}-1\big\}^{2}\big\{6-4(\frac{V_{0}}{V})^{\frac{2}{3}}\big\}\bigg]  
\end{align*}

where $E$ is energy, $V$ is volume, $B_{0}$ is equilibrium bulk modulus, $V_{0}$ is volume of experimental unit cell and $B_{0}^{\prime}$ is pressure derivative of bulk modulus at equilibrium value. The volume optimization process is carried out by varying the lattice parameters in a fixed ratio.

Fe$_{2}$VAl and Fe$_{2}$TiSn have the space group $Fm-3m$ and they are found to crystallise in  cubic $L2_{1}$ structure. Denoting these two full Heusler compounds as Fe$_{2}$YZ, where Y=V,Ti and Z=Al,Sn in the order, Fe atoms occupy the Wycoff position 8c $(\frac{1}{4},\frac{1}{4},\frac{1}{4})$, Y atoms occupy Wyckoff  position 4a $(0,0,0)$ and Z atoms occupy Wyckoff position 4b $(\frac{1}{2},\frac{1}{2},\frac{1}{2})$. We employed 5 different exchange-correlation functionals to study these compounds. The results are discussed in the sections below.

\section{Results and Discussion}
\subsection{\label{sec:level2}Structural properties evaluation}
In order to find the theoretical lattice constants of Fe$_{2}$VAl and Fe$_{2}$TiSn, calculations are carried out for several values of volumes corresponding to different lattice constants employing the five exchange-correlation potentials mentioned in section II. The obtained values of total energy is plotted as a function of volume. The Birch-Murnaghan(BM) parameters are used to fit the calculated data. The fitted curve gives the equilibrium lattice  constant and bulk modulus value. 5.762 {\AA}.\cite{web} and 6.074 {\AA}\cite{expta} are the reported values of experimental lattice constants for Fe$_{2}$VAl and Fe$_{2}$TiSn, respectively. These experimental values of lattice constants are used to construct the initial crystal structrure of the two compounds. The energy versus volume curves for the two compounds computed using five exchange-correlation functionals is shown in Fig. 1. 

\begin{figure*}
 
\begin{subfigure}{0.4\textwidth}
\includegraphics[width=0.85\linewidth, height=6.5cm]{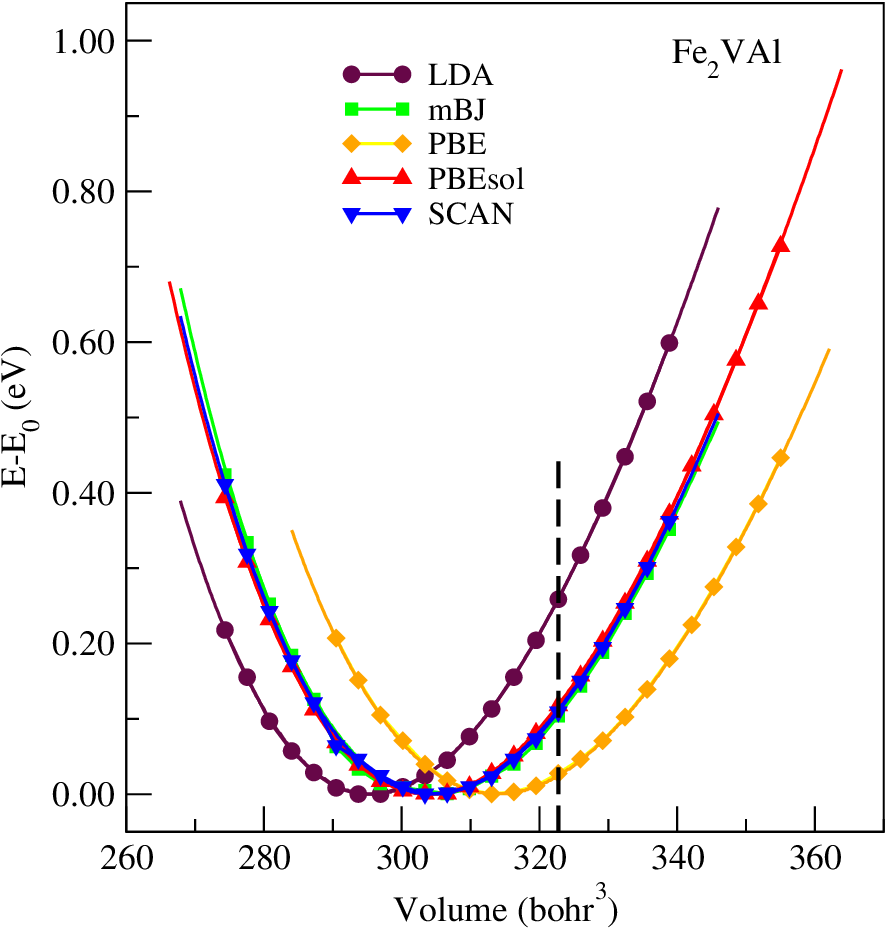} 
\caption{}
\label{fig:subim1}
\end{subfigure}
\hfill
\begin{subfigure}{0.4\textwidth}
\includegraphics[width=0.85\linewidth, height=6.5cm]{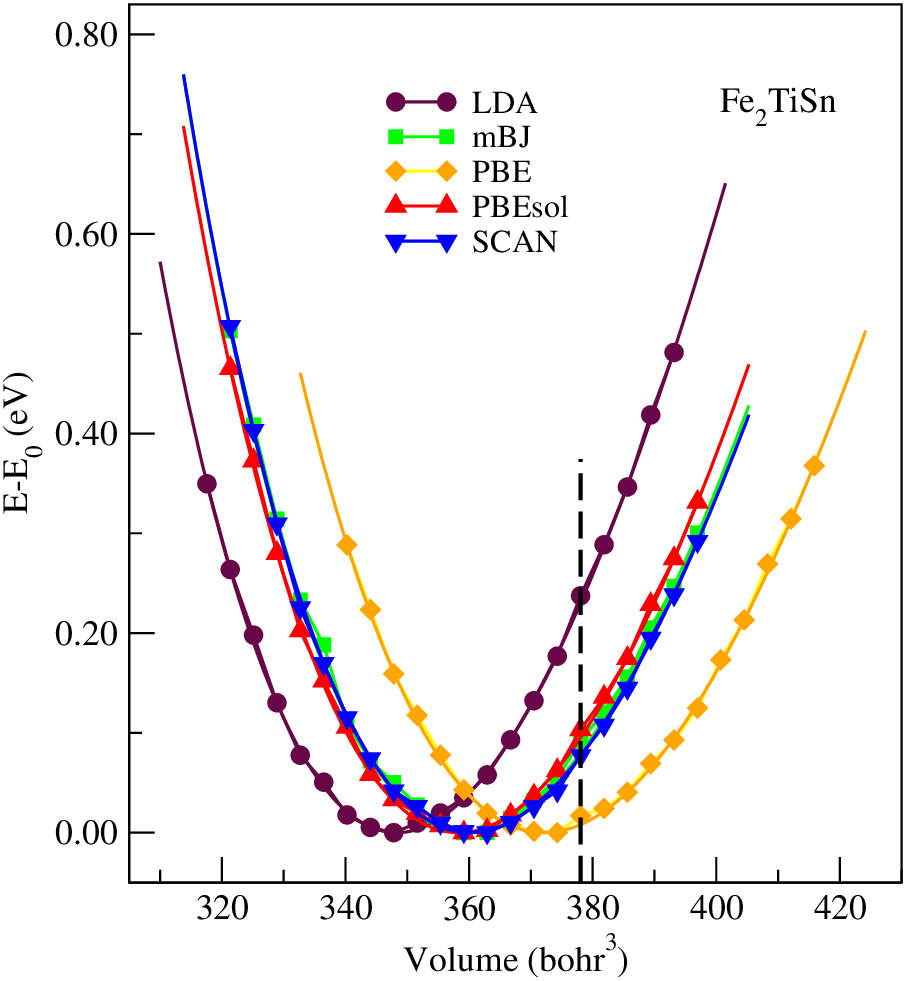}
\caption{}
\label{fig:subim2}
\end{subfigure}
 
\caption{Energy versus volume curves for (a)Fe$_{2}$VAl and (b) Fe$_{2}$TiSn\\}

\label{fig:image2}
\end{figure*}

In the figure, the symbols corresponds to the calculated values of the energy as a function of volume and the lines to the B-M fit to the calculated data, respectively. The dashed line perpendicular to the volume axis corresponds to the volume of experimental lattice constant. The energy axis is, $E-E_{0}$, the difference between volume dependent energy(E) and energy corresponding to equilibrium volume $E_{0}$. From the figure, the qualitative shifts in the calculated values from the experimental value can be made out. For the two compounds the large deviation in the lattice constants from LDA is noticeable. The PBE calculated values of optimized lattice constants lie close to the experimental lattice constant (nearer to the dashed line). For the other three potentials, the energy-volume parabolas lie in between PBE and LDA calculated values. Also, the values are close to each other.

The calculated values of optimized lattice constants $a_{o}$ and bulk modulus B$_{0}$ for the compounds are tabulated in Table.1 which support the behavior of the curves in Fig. 1. For Fe$_{2}$VAl, it is found from the table that LDA calculated value of lattice constant is the lowest of the remaining values and shows nearly 2.88 \% deviation from the experimental value. This clearly says LDA understimates the lattice constant. This value is 1.42 \% less than the value reported by Guo et al.($a_{o}=5.68 {\AA}$)using FPLAPW method.\cite{guo} The PBE calculated value is within the agreement of 1 \% and out of all 5 functionals showing fairly good agreement with experimental value. PBEsol, mBJ and SCAN are giving 1.98, 1.92 and 1.93 \% reduction from the experimental value, respectively. The PBEsol approximation,\cite{pbesol} which is specially constructed for the calculation of  lattice constant and it's dependent properties of solids and solid surfaces is found less suitable for full Heusler compounds. Because the calculated value of lattice constant for this compound from PBEsol functional is less than than that of the PBE, mBJ, SCAN calculated values.  

\begin{table*}
\caption{Calculated lattice constants $a_{0}$ and bulk modulus $B_{0}$ for the two compounds using five exchange-correlational functionals}
\resizebox{0.9\textwidth}{!}{%
\begin{tabular}{@{\extracolsep{\fill}}c c c c c c c c} 
 \hline\hline
 & \multicolumn{3}{c}{\textbf{Fe$_{2}$VAl}} & & \multicolumn{3}{c}{\textbf{Fe$_{2}$TiSn}}\\
 \cline{2-4} \cline{6-8}
       & Lattice constant $a_{0}$  & & Bulk modulus$B_{0}$ & & Lattice constant $a_{0}$& & Bulk modulus $B_{0}$\\
  & ({\AA}) & & (GPa) & & ({\AA})& & (GPa)\\
 \hline
LDA   & 5.5955 &   &266.96& &  5.9102 & &231.45\\
PBE   & 5.7089 & & 222.71 &  &6.0423 & &192.00\\
PBEsol&  5.6478 & & 246.05 & &5.9664 & &211.35\\
mBJ&  5.6512 & & 245.70 &  &5.9748& &207.83\\
SCAN&  5.6509  & & 249.40 &  &5.9762 & &205.49\\
 \hline\hline
\end{tabular}}
\end{table*}

For Fe$_{2}$TiSn, lattice constant calculated using LDA is 2.70 \% less than the experimental value. Also, LDA calculated value is the lowest of all the values found out using other functionals. PBE calculated value of lattice constant (6.0423 {\AA}) is close to the experimental value with deviation of only 0.52 \%. For both the compounds, it can also be observed that the last two functionals in the table are producing nearly same values and also showing improvement over PBEsol calculated value with respect to lattice constant. The table confirms the suggestion given by Sun \textit{et. al}\cite{scan} that SCAN functional is improvement over LDA and PBEsol but our calculated value shows not over PBE for the full Heusler alloys as PBE functional is giving a very good value of lattice constant even though it is producing lowest value of bulk modulus. 

The values of the equilibrium bulk modulus calculated for these two compounds are also tabulated in Table.1. 
It is obvious from the table, the overestimation in the value of $B_{0}$ by LDA and underestimation of the same by PBE. This observation is true for both Fe$_{2}$VAl and Fe$_{2}$TiSn. While in case of both the Heusler compounds the value of $B_{0}$ for PBEsol, mBJ and SCAN are lying close to each other. Kanchana \textit{et. al}(FP-LMTO method)\cite{kanchana}, and Hsu \textit{et. al}(FPLAPW method)\cite{hsubulk}. using PBE have reported the value of bulk modulus $B_{0}$, for Fe$_{2}$VAl to be 220.8 GPa, and 212  GPa, respectively. Our calculated value of 222.71 GPa using PBE functional is in near agreement with the values that calculated by Hsu \textit{et. al} and Sharma et al. For Fe$_{2}$TiSn, LDA obtained value of bulk modulus is 231.45 GPa and from PBE it is 192 GPa.
 
Here we have reported bulk modulus for for three new functionals. It can be observed that as in the case of lattice constant, the $B_{0}$ values from the last three functionals in the table for both the compounds are lying close to each other.  The experimental value of bulk modulus for these two compounds are not yet available for comparison. Thus, by observing the trend in lattice constant and bulk modulus values of these two full Heusler compounds, we can say that the relatively new functionals mBJ and SCAN are nearly equivalent approximations for full Heusler compounds for structural properties evaluation. The reason for overestimation of $B_{0}$ by LDA and underestimation by PBE functionals is as follows: We know that bulk modulus is calculated by the formula $B = V(\partial^2 E/\partial V^2)$. Where, V is volume of the unit cell and E is energy per unit cell. The term $(\partial^2 E/\partial V^2)$ in the equation gives the curvature of the energy versus volume parabolas of Fig. 1. In the figure, the curvature of LDA calculated curve is more than that of the PBE curve for the two compounds. Applying the values of the curvature and volume in the relation of the bulk modulus, clearly tells the reason behind this behavior.

\subsection{\label{sec:level2}Electronic structure analysis}
To know the behavior of the five exchange-correlational potentials in explaining the non-magnetic ground state  properties of the full Heusler compounds, we have carried out electronic structure analysis. Using the optimized lattice parameters dispersion curves and density of states are computed for each compound employing all the five exchange-correlation functionals. The dispersion curves calculated along the high symmetric k-points in the first Brillouin zone for selected functionals for the two Heusler compounds are shown in figure 2. The first and second row of the figure represents the dispersion curves for Fe$_{2}$VAl and Fe$_{2}$TiSn, respectively. 

\begin{figure*}
 
\begin{subfigure}{0.3\textwidth}
\includegraphics[width=0.8\linewidth, height=6.2cm]{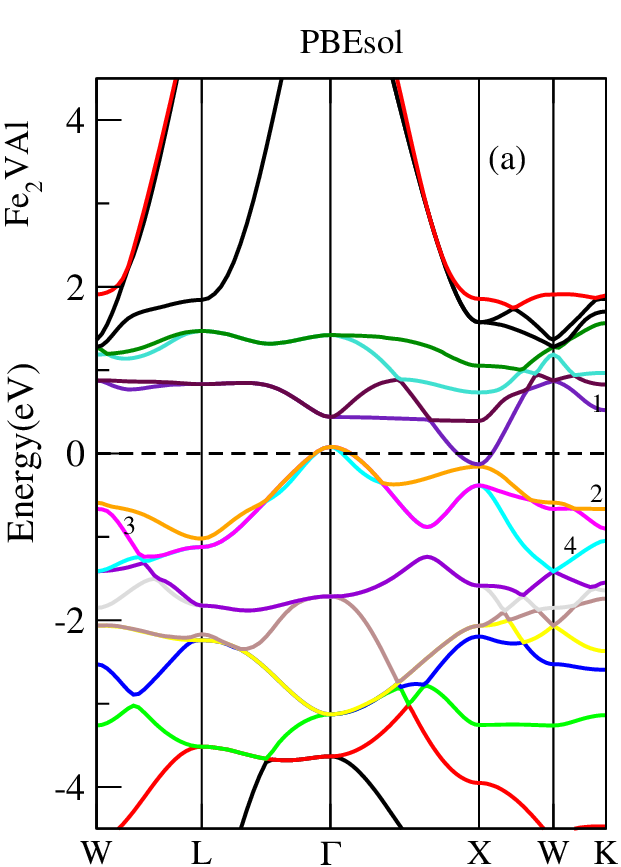} 
\end{subfigure}
\begin{subfigure}{0.3\textwidth}
\includegraphics[width=0.8\linewidth, height=6cm]{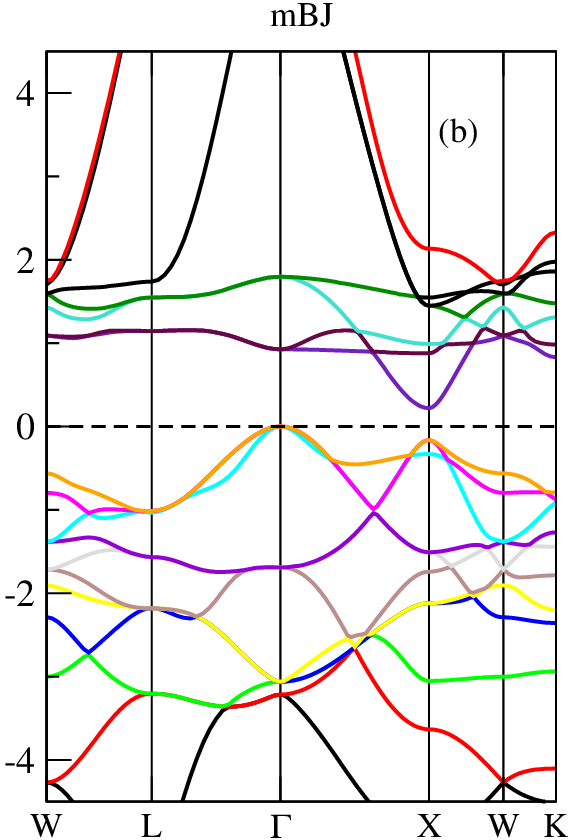}
\end{subfigure}
\begin{subfigure}{0.3\textwidth}
\includegraphics[width=0.8\linewidth, height=6cm]{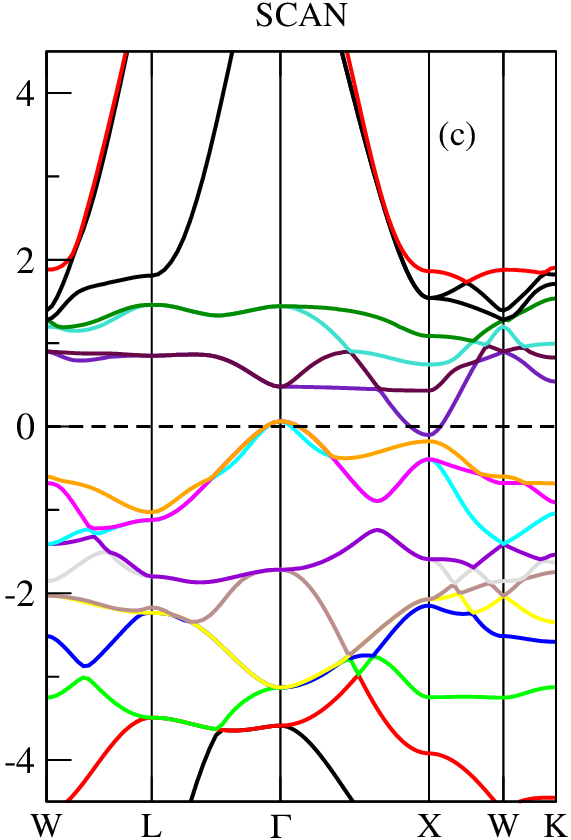}
\end{subfigure}

\vspace{0.8cm}

\begin{subfigure}{0.3\textwidth}
\includegraphics[width=0.8\linewidth, height=6cm]{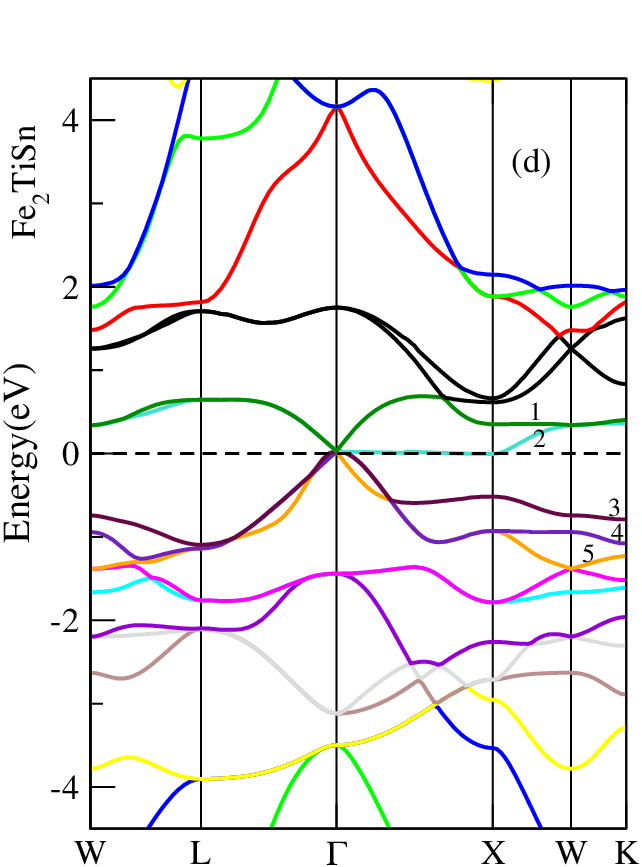}
\end{subfigure}
\begin{subfigure}{0.3\textwidth}
\includegraphics[width=0.8\linewidth, height=6cm]{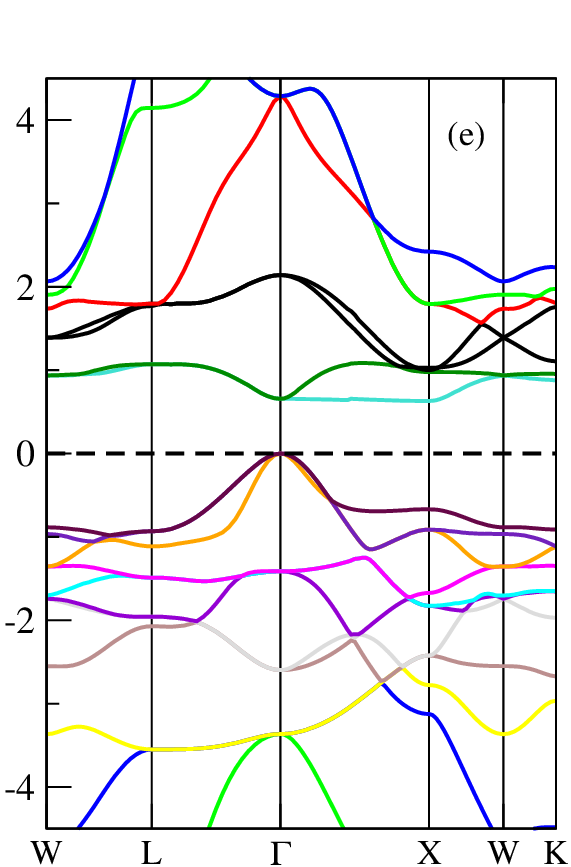}
\end{subfigure}
\begin{subfigure}{0.3\textwidth}
\includegraphics[width=0.8\linewidth, height=5.4cm]{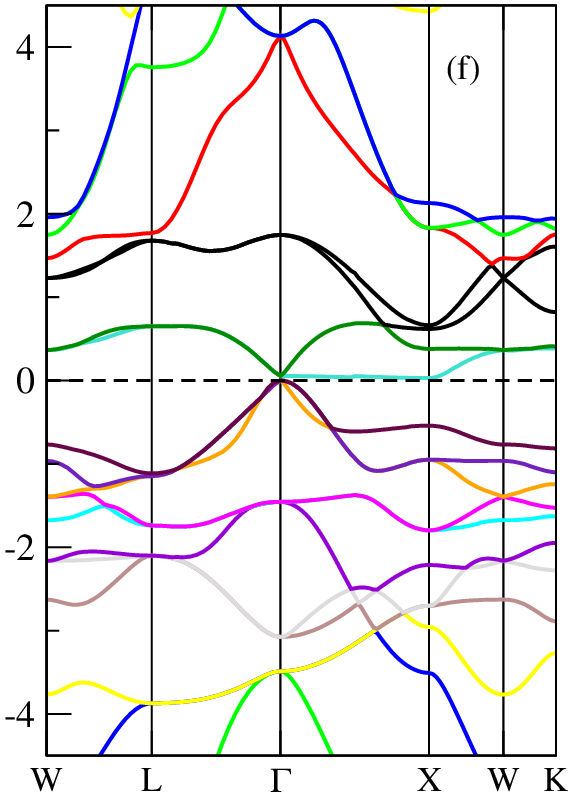}
\end{subfigure} 
\caption{Dispersion curves of Fe$_{2}$VAl and Fe$_{2}$TiSn from (a)\&(d)PBEsol, (b)\&(e)mBJ, (c)\&(f) SCAN functionals, respectively. }
\label{fig:image2}
\end{figure*}

The Fig. 2(a) shows the dispersion curves for Fe$_{2}$VAl obtained using PBEsol functional. The conduction band(CB) bottom at the X-point is crossing the $E_{F}$ and is lower in energy than the valence band(VB) at the $\Gamma$-point. This suggests the compound is semimetallic in nature. The direct gap is $\sim$0.36 eV above the VB maximum at the $\Gamma$-point. Also, the CB minimum is very close to the second VB maximum with energy gap of $\sim$20 meV. The negative band gap is defined as, $E_{g}=E_{c,min}-E_{v,max}<0$, where $E_{c,min}$ and $E_{v,max}$ are conduction and valence band extrema, respectively.\cite{mahanti} The value of the negative gap(or pseudo gap) from PBEsol calculation is -0.20 eV. The curves are triply degenerate at the top of VB. Along the $\Gamma$-X direction degeneracy is lifted and bands become doubly degenerate and non-degenerate at the X-point. Similary along the $\Gamma$-L direction. Also, bands are doubly degenerate at the CB bottom at $\Gamma$-point. This degeneracy is lifted from $\Gamma$ to X-direction. But the degeneracy is maintained along $\Gamma$-L direction. In case of LDA computed band structure of the compound, similar behavior is observed(not shown in the figure). The observed negative band gap is -0.24 eV from LDA calculation, implying the semimetallic nature. Thus, LDA and PBEsol are giving similar results for band structure. Not much improvement in PBEsol functional over LDA is found. LDA and PBEsol are known to underestimate the bandgap.\cite{bandgaplda,bandgapsol} 
In Fig. 2(c), the dispersion curves of Fe$_{2}$VAl calculated from SCAN exchange-correlation functional is presented. There is a formation of pseudogap by the overlap of VB maximum at the $\Gamma$-point and CB minimum at the X-point. The value of the pseudogap in this case is -0.16 eV. The direct gap above the VB maximum at $\Gamma$-point is 0.41 eV. The calculation using PBE is giving pseudo band gap of -0.13 eV which is nearer to that obtained from SCAN functional. The general features of the bands of the PBE(not shown in figure) are similar to that of the SCAN. Thus, both the dispersion curves calculated using PBE and SCAN predicting semimetallic nature in Fe$_{2}$VAl. 
The dispersion curve for Fe$_{2}$VAl is also calculated using mBJ potential which is constructed for the accurate bandgaps of semiconductors and insulators.\cite{mbj} The dispersion curve is shown in Fig. 2(b). The VB edge is at $\Gamma$-point and the CB bottom is the X-point. The band gap of 0.22 eV is observed which is indirect. Thus mBJ is predicting the compound to be semiconductor in nature. The enhancement of the band gap by the mBJ calculations compared to the other functional calculated gaps is clearly seen. There are noticeable changes in the characteristic of the mBJ calculated bands compared to that of other four functionals. The top of the valence band at X-point is doubly degenerate which is non-degenerate in case of dispersion curves of other four functionals. The direct gap above the $\Gamma$-point between valence band top and second conduction band bottom is 0.91 eV. Okamura \textit{et. al} have reported a bandgap of 0.1-0.2 eV for Fe$_{2}$VAl from photoconductivity measurements.\cite{okamura} The bandgap yielded by the mBJ calculation is in quite good agreement with this experimental band gap. 
The overall features of the band structure obtained from PBEsol and SCAN are almost the same with the difference in the values of overlap in the extrema of conduction and valence bands mentioned earlier. Thus we can say bands are shifted in energy from PBEsol to SCAN calculations.
\begin{figure*}
 
\begin{subfigure}{0.4\textwidth}
\includegraphics[width=0.9\linewidth, height=5cm]{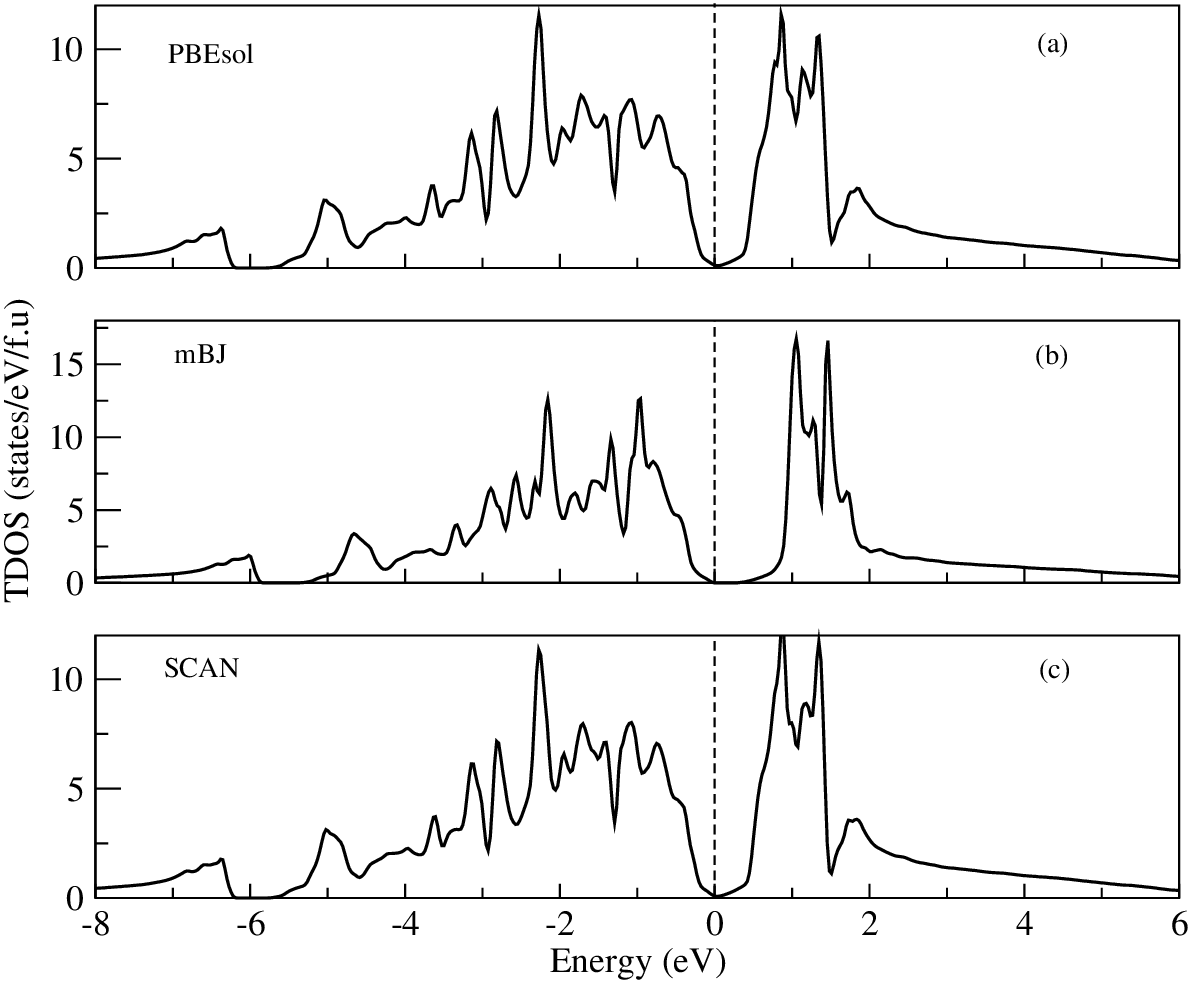} 
\end{subfigure}
\begin{subfigure}{0.4\textwidth}
\includegraphics[width=0.9\linewidth, height=5cm]{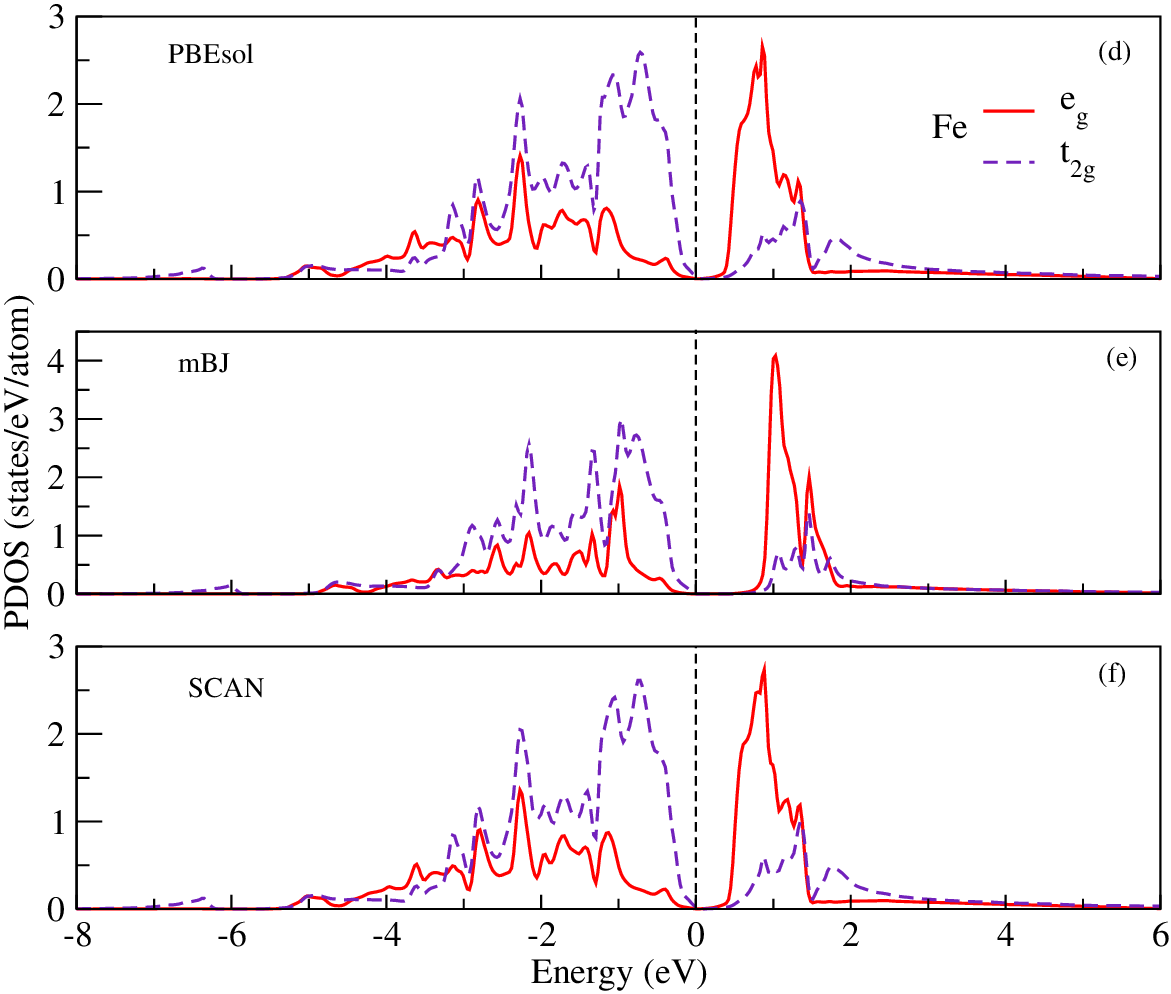}
\end{subfigure}

\vspace{0.9cm}
\begin{subfigure}{0.4\textwidth}
\includegraphics[width=0.9\linewidth, height=5cm]{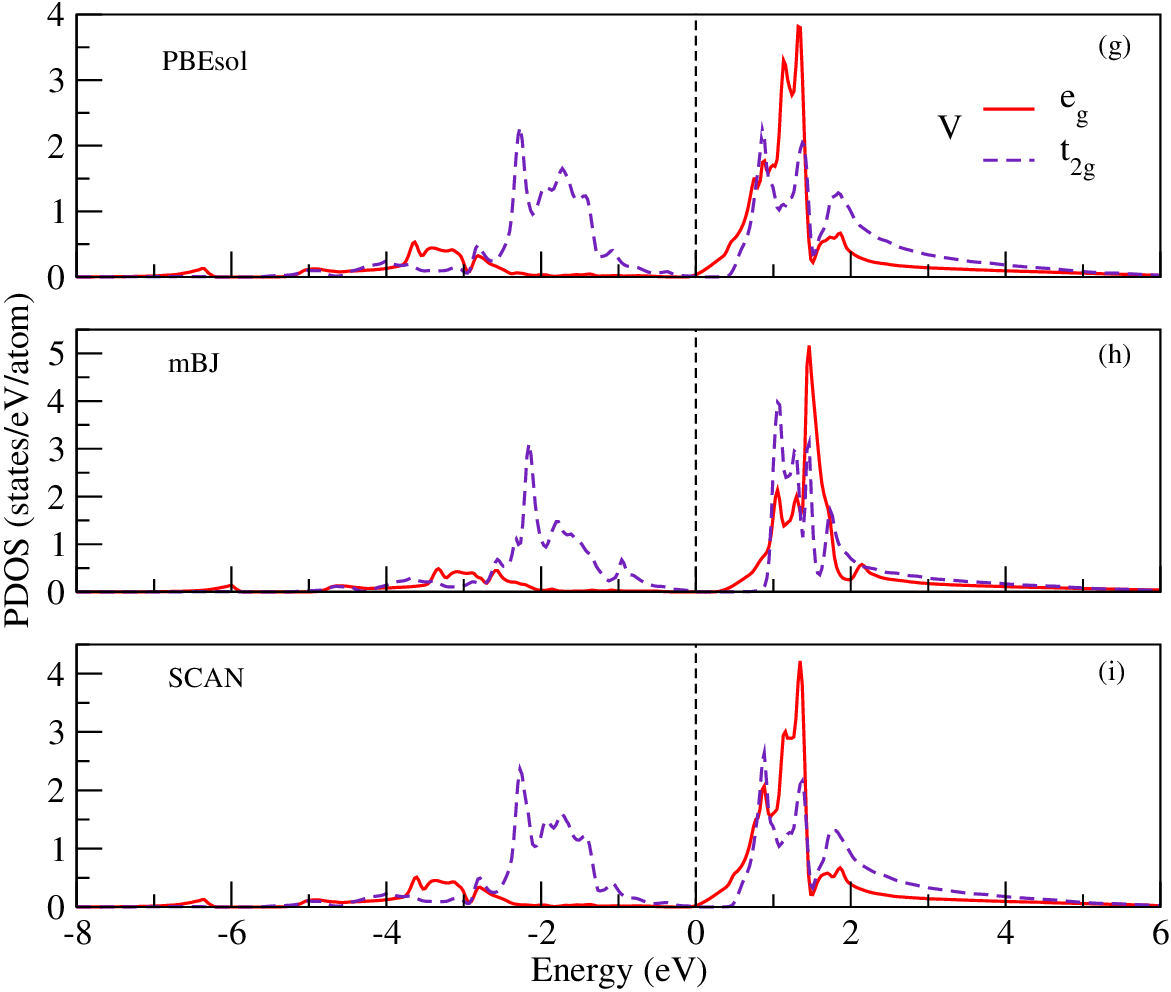}
\end{subfigure}
\begin{subfigure}{0.4\textwidth}
\includegraphics[width=0.9\linewidth, height=5cm]{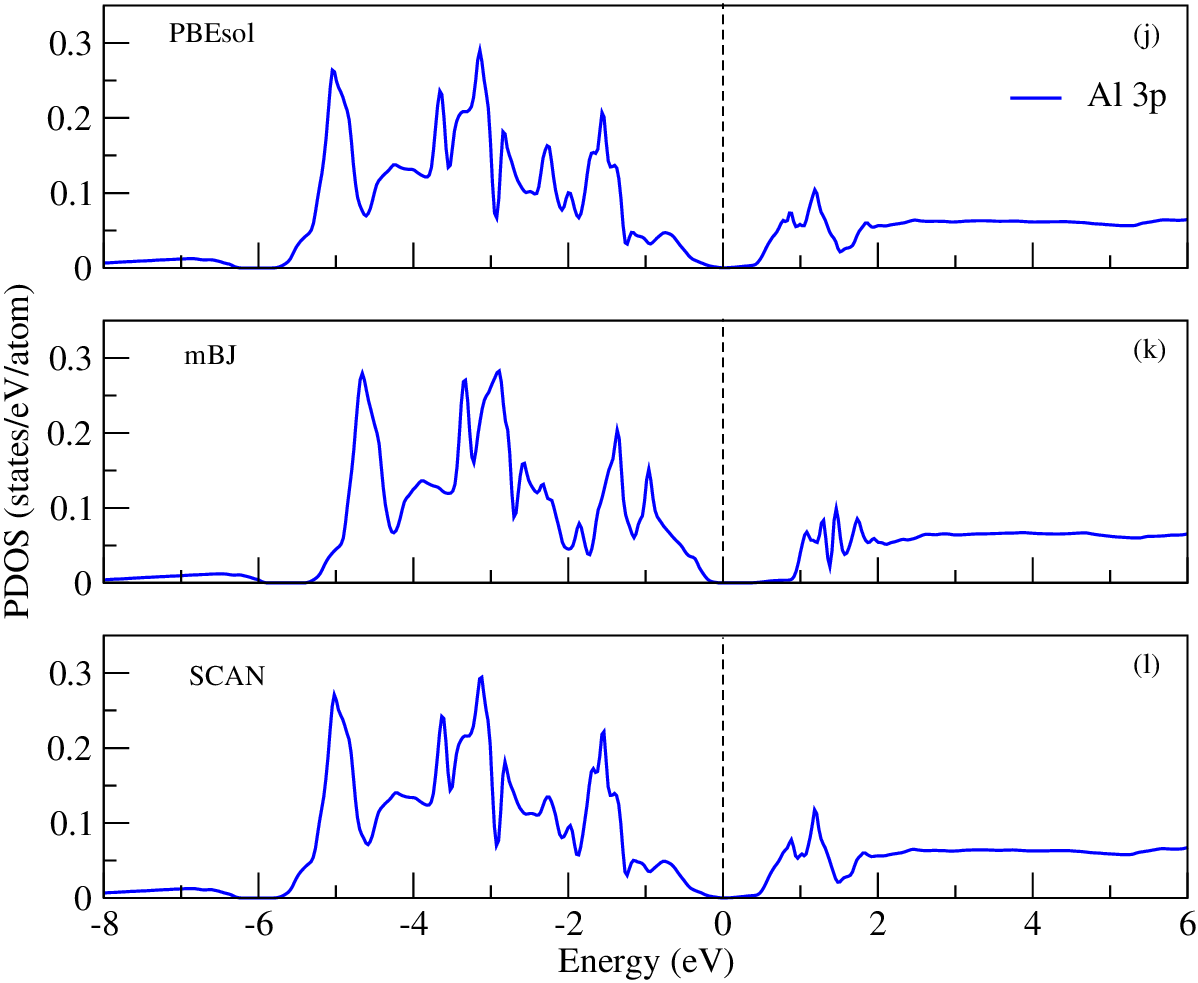}
\end{subfigure}
\caption{Total and partial density of states plots for Fe$_{2}$VAl obtained from PBEsol, mBJ and SCAN functionals. In (a)-(c) TDOS plots, (d)-(f) PDOS of Fe atom, (g)-(i) PDOS of V atom, (j)-(l) PDOS of Al atom are shown for three functionals}
\label{fig:image2}
\end{figure*}

Fig. 2(d)\&(f) shows the dispersion curves calculated for Fe$_{2}$TiSn using PBEsol and SCAN functionals. The general features of the band structures of the compounds are: The valence band top at the $\Gamma$-point is triply degenerate and the conduction band bottom at $\Gamma$-point is doubly degenerate. The bands are degenerate along the $\Gamma$-L direction and non-degenerate along the $\Gamma$-X direction near the Fermi energy in the conduction band region. There is a flat conduction band present along the $\Gamma$-X direction in conduction band region. The presence of the flat conduction band results in higher value of effective mass which makes the compound good for thermoelectric application. In the case of PBEsol and LDA obtained dispersion curves(not included in figure) there is direct overlap of the conduction band and valence band at the $\Gamma$-point. Also, the bottom of the CB at the X-point is crossing the $E_{F}$. This behavior may be because of the understimation of the band gap by LDA and PBEsol functionals. These two functionals are suggesting  semi-metallic type of behavior of the compound. In case of SCAN and PBE functionals, the dispersion curves showing the CB minimum at the X-point and the VB maximum at the $\Gamma$-point(for PBE not shown in figure). This indicates the compound is an indirect band gap semiconductor. The indirect band gap yielded in case of SCAN functional is 0.028 eV and that from the PBE functional is 0.033 eV. The direct gap at the $\Gamma$-point in Fig. 2(f) is 0.052 eV. While, PBE is showing the direct gap of 0.056 eV. Thus these values and nature of the dispersion curves suggest that the functionals SCAN and PBE are behaving in similar way. While, PBEsol and LDA functionals are yielding similar results for Fe$_{2}$TiSn.
 
The mBJ potential is found to enhance the band gap. The enhanced indirect band gap for the compound compared to other two functionas is seen from Fig. 2(e). The indirect band gap value obtained from mBJ is 0.68 eV. We have not come across any experimental band gap value for Fe$_{2}$TiSn Heusler compound.
Thus, PBE, SCAN and mBJ are found to open the band gap with different shift in energy to a higher value while, PBEsol and LDA are underestimating the gap producing zero gap for Fe$_{2}$TiSn.

	For Fe$_{2}$VAl Heusler compound, Bilc \textit{et. al} used B1-WC hybrid functional to study the electronic and thermoelectric properties and reported an indirect band gap of $\sim$1 eV. They claimed B1-WC hybrid functional is accurate for the calculation of electronic and thermoelectric properties.\cite{bilc15} However, this gap is $sim$0.78 eV higher than the mBJ produced value as well as the experimental value. It is important to note that thermoelectric properties are sensitive to the temperature dependent band gap value and hence, the description of the thermoelectric behavior of the compound using this approach may not be correct. Also, for Fe$_{2}$TiSn , the same group reported a band gap of $\sim$1 eV using the hybrid functional\cite{bilc15}, which in case of mBJ obtained value of ours is 0.68 eV.   Meinert, Markus using mBJ potential and FPLAPW method obtained a band gap value of 0.31 eV for Fe$_{2}$VAl and 0.69 eV for Fe$_{2}$TiSn, respectively.

For the compounds Fe$_{2}$VAl and Fe$_{2}$TiSn, total density of states(TDOS) and partial density of states(PDOS) are calculated using five exchange-correlational functionals. Fig. 3 shows the TDOS and PDOS for Fe$_{2}$VAl calculated using PBEsol, mBJ and SCAN functionals. In Fig. 3(a)-(c) the TDOS are depicted. PBEsol and SCAN functionals are yielding pseudogap in the vicinity of Fermi level. Near the Fermi level sharp peaks of DOS are seen from all three calculations and the presence of pseudogap is yielding a minute DOS at the Fermi level ($E_{F}$) . In the figure the dotted line at zero energy represents the Fermi level. In case of mBJ obtained TDOS plot a large band gap is clearly seen with zero density of states present at the $E_{F}$. The TDOS plot of mBJ says Fe$_{2}$VAl is a semiconductor with band gap of 0.22 eV as predicted by dispersion curves. In order to understand the contribution from constituent atoms, PDOS are calculated and are shown in Fig.3 for Fe, V, and Al atoms. For Fe and V atoms the three fold degenerate states($d_{xy}, d_{yz}, d_{zx}$) are represented by t$_{2g}$ and two fold degenerate orbitals ($d_{x^{2}-y^{2}}, d_{3z^{2}-r^{2}}$) are represented by e$_{g}$ states. It is clearly seen from the figure that near the Fermi level in the conduction band region Fe t$_{2g}$ states(Fig. 3(d)-(f)) are main contributors in the range -1.2 to 0 eV. The same behavior is observed for all three functionals except from mBJ, the number of states contribution is more. From -2.5 to -1.2 eV both V and Fe t$_{2g}$ states are contributing to the conduction band.

The V e$_{g}$ states are contributing less to the valence band region as seen in Fig. 3(g)-(i). While contribution from Fe e$_{g}$ states to valence band are more compared to V e$_{g}$ states. In the range 0 to 2 eV high intensity peaks of V and Fe e$_{g}$ states are observed. In the 0-2 eV region the intensity of t$_{2g}$ states of V atoms are more compared to that of Fe atoms. This indicates that valence band top is more of  Fe t$_{2g}$ character and bottom of conduction band region is more of V e$_{g}$ character. The formation of pseudogap is due to the overlap of these two bands in the vicinity of Fermi level. The contribution to the DOS from Al 3p orbitals in case of three funcitonals are shown in Fig. 3(j)-(l). The intensity of the peaks of Al 3p states are high in the valnce band region in -6 to -1 eV. The intensity of the PDOS peaks calculated by mBJ functional for V and Fe atoms are more compared to that of other functionals. But in case of Al atom mBJ and other functionals giving nearly same contribution to PDOS. The obtained DOS from LDA and PBE functionals(not shown in figure) are in near agreement with PBEsol and SCAN characterization of electronic states, respectively.

The total density of states(TDOS) and partial density of states(PDOS) calculated for Fe$_{2}$TiSn Heusler compound using three  exchange-correlation functionals are shown in Fig. 4. The Fig. 4(a)-(c) represents the TDOS plots from PBEsol, mBJ and SCAN functionals calculations for Fe$_{2}$TiSn. From the TDOS plots of Fe$_{2}$TiSn, it can observed that Fe$_{2}$TiSn is semiconductor in nature. The width of gap is enhanced and the value of the gap is more in case of mBJ calculations. The states are shifted by $\sim$ 0.5 eV higher in energy in the conduction band region in case of mBJ obtained DOS(Fig. 4(b)). Fig. 4(d)-(f) and (g)-(i) show the PDOS of Fe and Ti atoms calculated from three functionals. Near the Fermi level, the valence band region from -1 to -0.5 eV is mostly of Fe t$_{2g}$ character. The small but finite density of states at the Fermi level is due to the overlap from the Fe t$_{2g}$ orbitals. It is clear from these figures that the contribution to conduction band is mainly from the e$_{g}$ states of Fe and Ti atoms. But, mBJ functional is showing equal contribution to the conduction band region from Ti t$_{2g}$ states. The three functionals suggest the contribution to the valence band region is mostly from t$_{2g}$ states of both Fe and Ti atoms. In the lower energy region of -6 to -2 eV the contribution to DOS is from the Sn 5p orbitals(Fig. 4(j)-(l)). Also, from the TDOS plots it is observed that the value of TDOS produced by mBJ potential at the higher energy region of valence band is relatively high to that of other functionals.

These observations lead to an understanding that, for the two compounds, mBJ approximates the ground state electronic structure with a notable difference from the other functionals. For the two compounds, mBJ is opening a reasonable gap compared to other functionals, with the band gap value matching to experimental value in case of Fe$_{2}$VAl. The effective mass is a property dependent on the shape of the bands, which is important in explaining transport properties. Thus band structure is important in explaining the carrier dynamics in different energy levels. DOS plots show sharp peaks near the Fermi level, which is a needed feature for a thermoelectric material.\cite{mahansofo} 
\begin{figure*}
 
\begin{subfigure}{0.4\textwidth}
\includegraphics[width=0.9\linewidth, height=5cm]{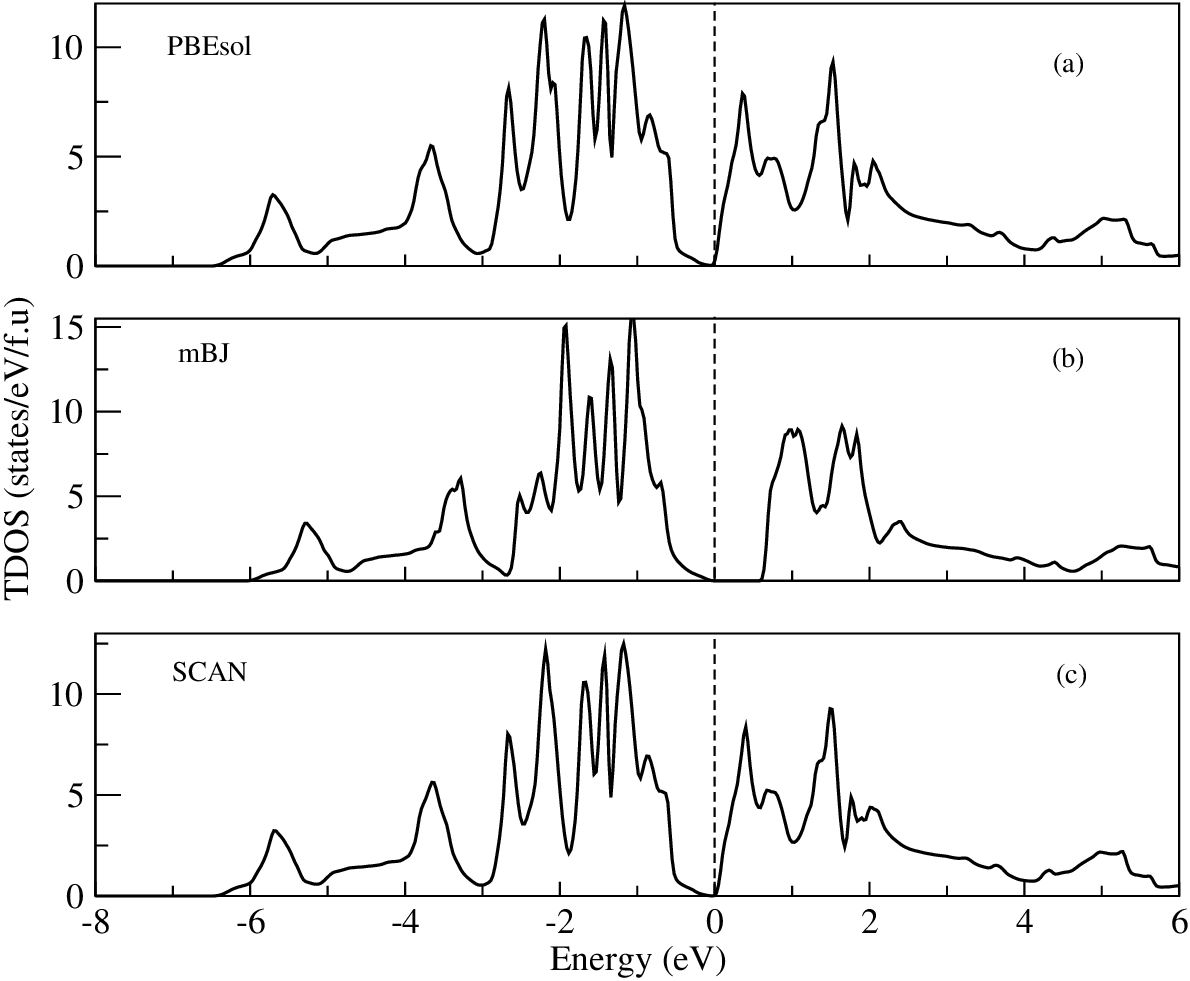} 
\end{subfigure}
\begin{subfigure}{0.4\textwidth}
\includegraphics[width=0.9\linewidth, height=5cm]{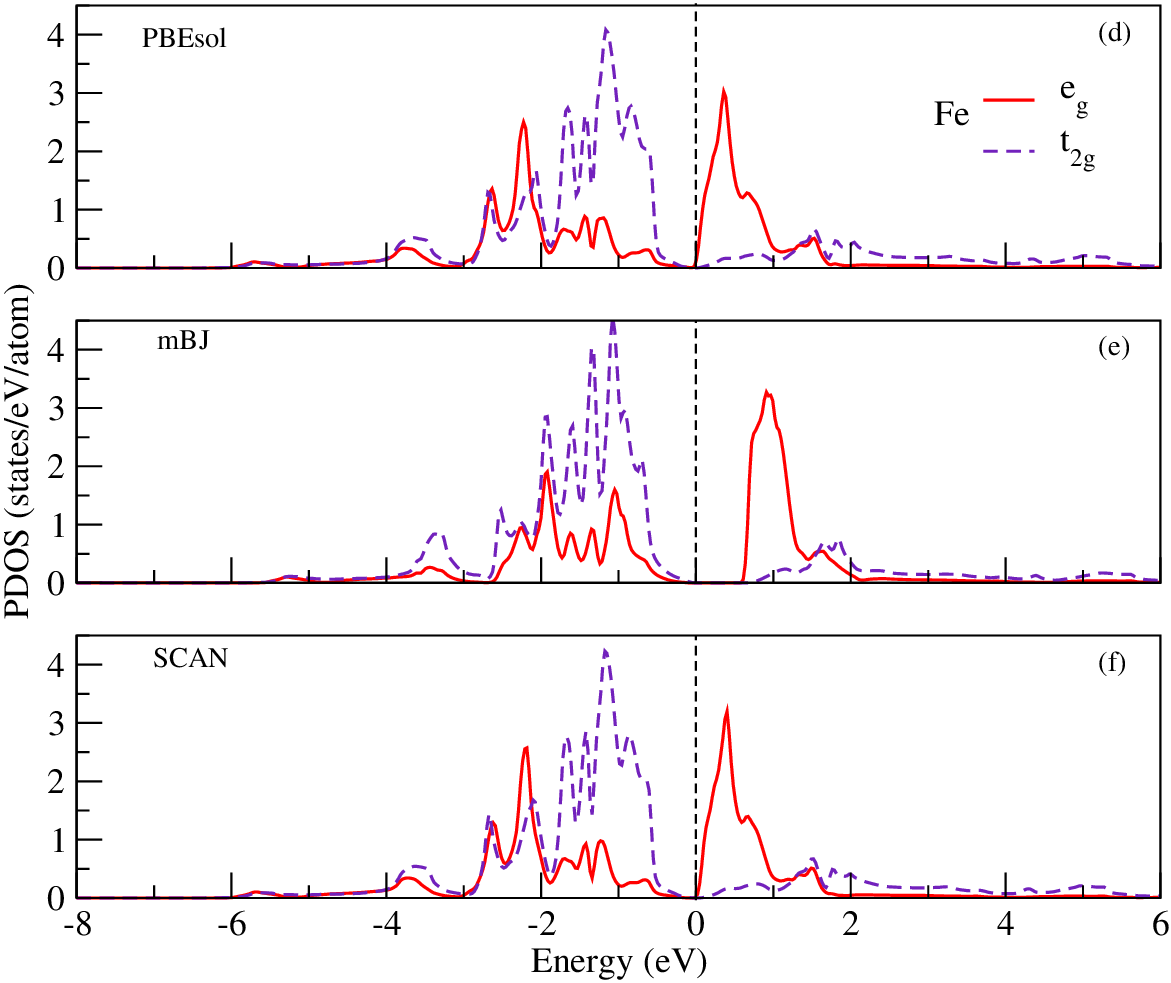}
\end{subfigure}

\vspace{0.9cm}
\begin{subfigure}{0.4\textwidth}
\includegraphics[width=0.9\linewidth, height=5cm]{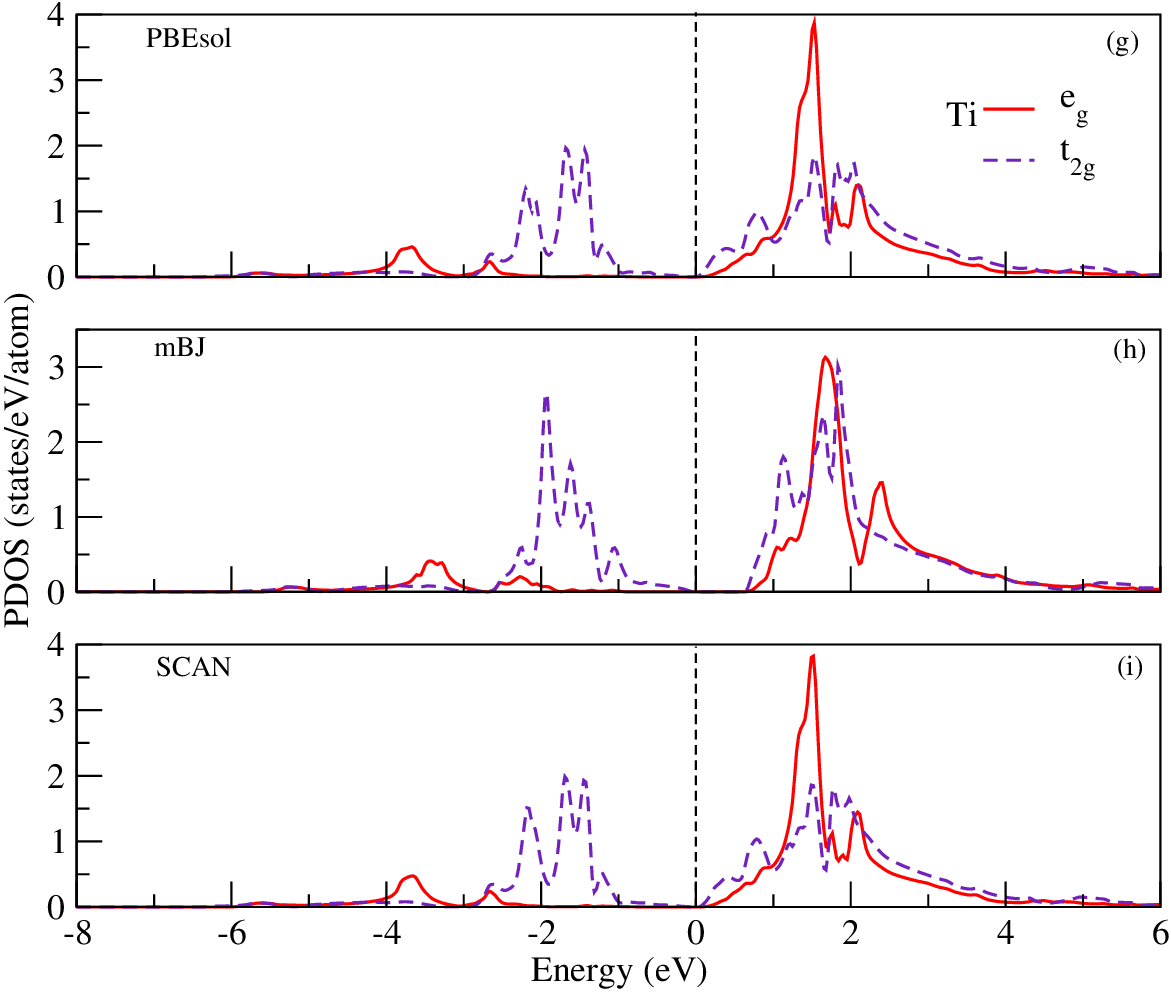}
\end{subfigure}
\begin{subfigure}{0.4\textwidth}
\includegraphics[width=0.9\linewidth, height=5cm]{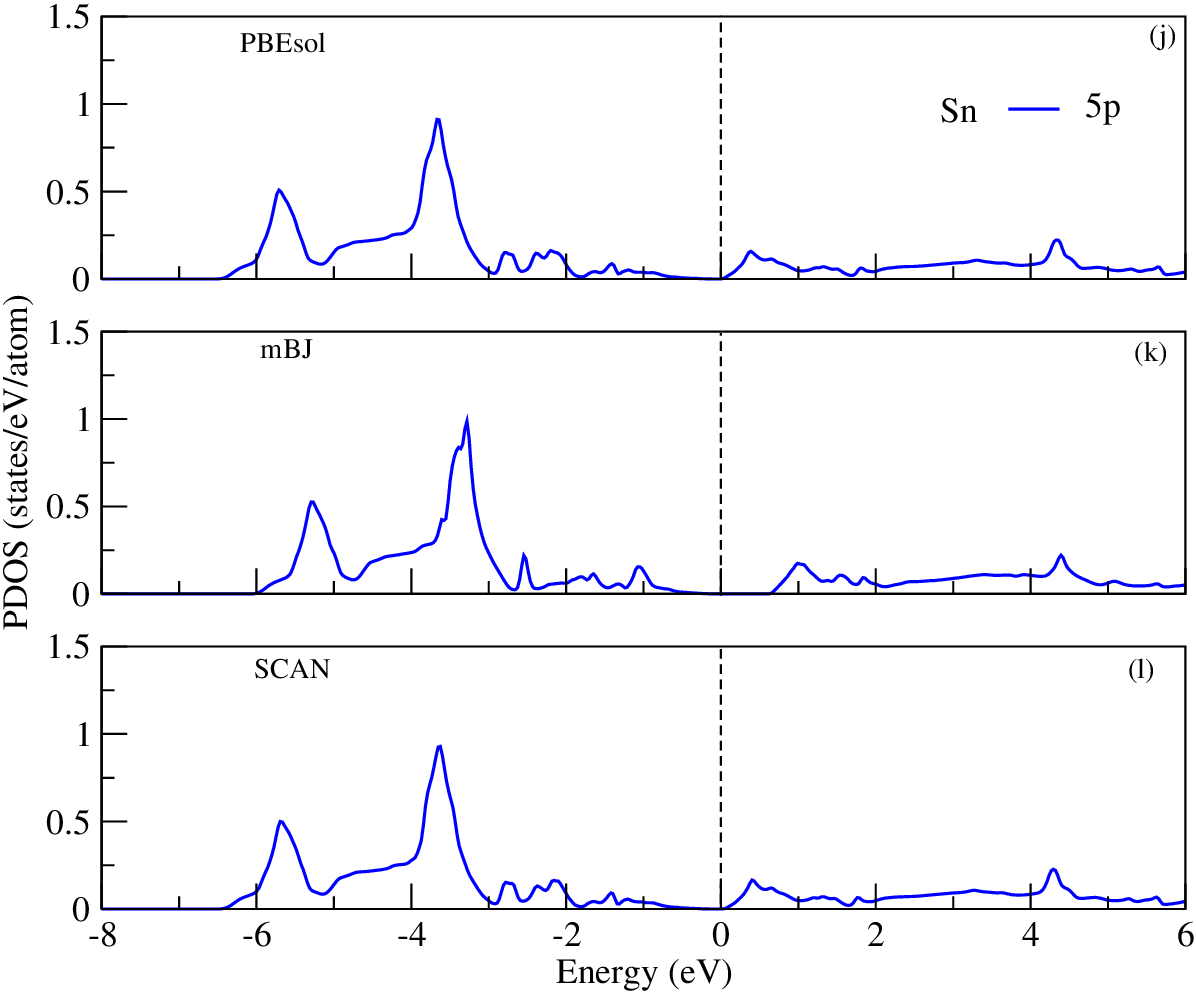}
\end{subfigure}
\caption{Total and partial density of states plots for Fe$_{2}$TiSn obtained from PBEsol, mBJ and SCAN functionals. In (a)-(c) TDOS plots, (d)-(f) PDOS of Fe atom, (g)-(i) PDOS of Ti atom, (j)-(l) PDOS of Sn atom are shown for three functionals}
\label{fig:image2}
\end{figure*}

	To understand, how using different exchange-correlation functionals to study single compound affects the effective mass, we have calculated effective mass($m^{*}$) of charge carriers(holes and electrons) along the high symmetric directions in the first Brillouin zone.  The bands that contribute significantly to the transport properties in Fe$_{2}$VAl and Fe$_{2}$TiSn are labelled with numbers in Fig. 2(a)\&(d). Effective mass ($m^{*}$) is calculated for  charge carriers in these bands and expressed in terms of electrons mass($m_{e}$). The calculated $m^{*}$ for Fe$_{2}$VAl and Fe$_{2}$TiSn compounds are tabulated in Table 2 \& 3, respectively. In the table, for instance, $\Gamma$-$\Gamma$X denotes the effective mass calculated at  $\Gamma$-point along $\Gamma$-X direction. Similar meaning is conveyed in other high symmetric directions. The symbols B1,B2,..B5 stand for bands numbered 1,2,..5, respectively. 

\begin{table*}
\caption{Effective mass($m^{*}$) of charge carriers calculated for Fe$_{2}$VAl compounds at $\Gamma$ and X points, respectively in various bands.}

\scalebox{1.15}{
\resizebox{0.8\textwidth}{!}{%
\begin{tabular}{@{\extracolsep{\fill}} c c c c c c c c c c c c c c c c c c c c c c c c c}
 \hline\hline

 & \multicolumn{3}{c}{LDA} & & \multicolumn{4}{c}{PBE} & & \multicolumn{4}{c}{PBEsol} & & \multicolumn{4}{c}{mBJ} & & \multicolumn{4}{c}{SCAN}\\
 \cline{2-4} \cline{6-9} \cline{11-14} \cline{16-19} \cline{21-24}
        &B2  &B3 &B4 &  &B1 &B2  &B3 &B4 & &B1 &B2  &B3 &B4 & &B1 &B2  &B3 &B4 & &B1 &B2  &B3 &B4\\
 \hline
$\Gamma$-$\Gamma$X    &1.08 &1.06 &0.37 &  &-- &1.19 &1.13 &0.40 &  &-- &1.14 &1.07 &0.39 &  &-- &1.22 &1.19 &0.64 &  &-- &1.15 &1.11 &0.40\\
$\Gamma$-$\Gamma$L    &0.64 &0.59 &0.58 &  &-- &1.39 &0.73 &0.60 &  &-- &1.19 &0.62 &0.60 &  &-- &1.29 &1.20 &0.59 &  &-- &1.20 &0.73 &0.61\\
X-XW   &1.09 &-- &-- &  &0.42 &0.96 &-- &-- &  &0.37 &0.92 &-- &-- &  &0.22 &0.34 &0.25 &1.05 &  &0.36 &0.93 &-- &--\\
X-X$\Gamma$    &4.28 &-- &-- &  &0.77 &3.27 &-- &-- &  &0.82 &3.05 &-- &-- &  &0.71 &0.37 &0.30 &4.48 &  &0.84 &3.17 &-- &--\\

 \hline
 
\end{tabular}}}
\end{table*}

\begin{table*}
\caption{Effective mass($m^{*}$) of charge carriers calculated for Fe$_{2}$TiSn compound at $\Gamma$ and X points respectively, in various bands.}
\scalebox{1.1}{
\resizebox{0.7\textwidth}{!}{%
\begin{tabular}{@{\extracolsep{\fill}} c c c c c c c c c c c c c c c c c c c}
 \hline\hline

 & \multicolumn{2}{c}{LDA} & & \multicolumn{2}{c}{PBE} & & \multicolumn{2}{c}{PBEsol} & & \multicolumn{5}{c}{mBJ} & & \multicolumn{2}{c}{SCAN}\\
 \cline{2-3} \cline{5-6} \cline{8-9} \cline{11-15} \cline{17-18}
      &B2  &B3 &   &B2  &B3  &  &B2  &B3  & &B1 &B2  &B3 &B4 & B5 &  &B2  &B3\\
 \hline
$\Gamma$-$\Gamma$X   &-- &0.63 &   &-- &0.64 &   &-- &0.64  &  &0.53 &-- &0.67 &0.69 &0.38 &   &50.20 &0.65 \\
$\Gamma$-$\Gamma$L    &-- &0.32  &   &-- &0.33  &   &-- &0.33  &  &0.99 &0.96 &0.72 &0.73 &0.33 &   &-- &0.33 \\
X-XW   &0.65 &--  &   &0.87 &--  &   &0.74 &--  &   &-- &0.96 &-- &-- &--  &   &0.79 &-- \\
X-X$\Gamma$    &36.36 &--  &   &37.12 &--  &   &36.45 &--  &  &-- &36.07 &-- &-- &-- &   &36.53 &-- \\
\hline
 
\end{tabular}}}
\end{table*}

The effective mass($m^{*}$) is calculated by the formula $m^{*}= \hbar^{2}/({d^2 E}/{d k^2})$ under parabolic approximation\cite{ashcroft}. According to this formula, the value of effective mass at a k-point is decided by the shape of the dispersion curve. This means, in Fe$_{2}$TiSn, large effective mass is expected because of the presence of a flat conduction band along $\Gamma$-X direction. In Fe$_{2}$VAl, the bands B1, B3, B4 are triply degenerate at $\Gamma$-point. At X-point B2, B3 are doubly degenerate in mBJ calculated dispersion curves(Fig. 2(b)) and in case of other functionals doubly degeneracy is changed to B3, B4 bands at lower energy position. Effective mass at these points are calculated by fitting the band edges with a parabola. While, in Fe$_{2}$TiSn, the bands B1, B2 are doubly degenerate and B3, B4 and B5 are triply degenerate at the $\Gamma$-point. The shape of the bands 4, 5 and 1 at X-point resembles closely of a cone. This feature of the bands is observed in case of LDA, PBEsol, PBE and SCAN functionals. Parabolic approximation cannot be applied for these cases to get the value of effective mass. But, the edges of degenerate bands at $\Gamma$-point of mBJ dispersion curves are parabolic in nature and fitted with parabola to obtain the value of the effective mass.

\subsection{\label{sec:level2}Thermoelectric properties}
We have studied the Fe$_{2}$VAl and Fe$_{2}$TiSn full Heusler compounds using five exchange-correlation functionals. It is well know that both the compounds are good thermoelectric materials. It is important to understand the electronic structure in order to explain the thermoelectric properties, since Seebeck coefficient($S$) is related to effective mass($m^{*}$) and carrier concentration(n). From the free electron theory approximation the relation between Seebeck coefficient and effective mass is given by the relation\cite{formula},
\begin{equation}
S = (8\Pi^{2}k^{2}_{B}/3eh^{2})m^{*}T(\Pi/3n)^{2/3}
\end{equation}
where, $k_{B}$ is Botlzmann constant, $e$ is electronic charge, $h$ is Planck's constant, and $n$ is carrier concentration respectively. The above equation says Seebeck coefficient is directly related to the value of the effective mass. Thus the calculation of the effective mass of charge carriers in the bands governing the transport properties is important. This can explain which bands are contributing more towards the Seebeck coefficient.
In case of Fe$_{2}$VAl, the value of the indirect band gap from the mBJ calculations is 0.22 eV which is in near agreement with the experimental band gap. The bands 2, 3, 4 at the top of the valence band(VB) at $\Gamma$-point are triply degenerate as discussed earlier, but contribution to the transport property will be different since the shapes of the bands are different (Fig. 2(b)). This can be confirmed from the Table 2. as the effective mass of band 4 at $\Gamma$-point along X and L directions is less than that of the bands 2 and 3. The carriers in bands B2 and B3 are main contributors to Seebeck coefficient with effective mass more than $m_{e}$(mass of electron) at $\Gamma$-point and X-point. Except for mBJ, the effective mass is more than 3 times at X-point along the $\Gamma$-direction. Qualitatively, also from the curvatures of the bands this fact can be understood. The VB top at X-point is doubly degenerate(bands 2 \& 3), and at the higher temperatures the fraction of electrons jumping across the direct gap of 0.39 eV will be more. So, effective mass at X-point is also calculated. The difference in energy between the 1st and 2nd VB at the X-point is 0.17 eV. This corresponds to the temperature of $\sim$2000 K. Therefore, there will be negligible contribution from this band(B4). The calculations from PBEsol, SCAN, LDA and PBE functionals does not show any real gap in the compound. Since the mBJ calculation is producing the experimental band gap, if we create a band gap artificially, equal to the mBJ obtained gap in band structures obtained from other functionals, now we can utilise the effective mass calculated from other functionals to explain the transport properties of Fe$_{2}$VAl. In the dispersion curves of the other functionals, as explained in previous section the features are different therefore, the effective mass computed for various bands is different from different functionals. The band gaps and position of bands at various k-points obtained after shifting the bands will be different. This implies the number of eletrons and holes created because of the transition between the bands now will not be same as before from all the functionals. Therefore, using this 	concept and the calculated values of effective mass one can find which functional can better explain the thermoelectric properties. 

	The effective mass calculated for Fe$_{2}$TiSn are in Table 3. The effective mass of holes along the X-$\Gamma$ direction is very high. This is due to the presence of the flat band along that direction. This means holes are the major contributors to Seebeck coefficient($S$) in case of Fe$_{2}$TiSn and high value of $S$ is expected in the compound. The high value of effective mass for the flat band(B2) from Table. 3 supports this argument, which is more than 36 times mass of electron along the X-$\Gamma$ direction. The indirect gap observed from mBJ calculation is 0.68 eV and from the SCAN(PBE) calculation 0.028eV(0.033eV). Thus, at higher temperatures the probability of occupation of the thermally excited electrons at the lowest vacant conduction band is higher in case of SCAN and PBE predicted properties compared to mBJ calculations. This means the Seebeck coeffiecient value approximated from these functionals should be less as Seebeck coefficient is inversly proportional to the carrier concentration. The experimental band gap for the compound is not yet known. Thus, mBJ should better explain the thermoelectric behavior of Fe$_{2}$TiSn, with higher value of band gap and large effective mass, if the bandgap value is matching with the experimental band gap. Thus, effective mass is an important quantity in the calculation of Seebeck coefficient and hence figure of merit of a thermoelectric.

\section{Conclusions} 
In this work, we have studied Fe$_{2}$VAl and Fe$_{2}$TiSn, two full Heusler alloys using five exchange-correlation(XC) functionals viz., LDA, PBE, PBEsol, mBJ and SCAN. The two compounds are experimentally studied compounds. Structural properties of both the compounds are evalulated using five density functionals. It is observed that, for both the compounds, bulk modulus is understimated by PBE and overestimated by LDA functional, while lattice constant is underestimated by LDA. Out of all five functionals PBE calculated value of lattice constant is nearer to the experimental lattice constant. PBEsol, mBJ and SCAN are giving values of lattice constant and bulk modulus in between the values that are from LDA and PBE functionals. To understand the electronic ground state properties for the two compounds dispersion curves and density of states are calculated. For Fe$_{2}$VAl, mBJ calculations yield an indirect band gap of 0.22 eV which is in reasonable agreement with the experimental value. While for Fe$_{2}$TiSn the value of indirect gap from the same functional is 0.68 eV. The general features of the dispersion curves observed from LDA and PBEsol calculations, and PBE and SCAN calculations are similar. While, mBJ features showing more changes with respect to other functionals for both the compounds. The LDA and PBEsol functionals are found to underestimate band gaps in the compounds. Effective mass of charge carriers are calculated from the dispersion curves applying parabolic approximation and contributions to transport properties are discussed. Very high value of effective mass for holes are obtained in case of Fe$_{2}$TiSn compound. For structural properties calculation, we found that SCAN functional is an improvement over LDA, PBE and PBEsol for these two compounds. 
If description of the any one of the five exchange-correlation functional is the most appropriate, then the transport properties calculated using that functional should be in well agreement with the experimental observations. We would like to investigate this aspect by calculating the transport coeffiecients and thermoelectric properties of the two Heusler compounds with non-magnetic ground state and thereby compare with the experimental values in our next work. Thus, this work is intended to give a hint on, which functional gives a better description of electronic structure of full Heusler compounds for thermoelectric applications.
 
\section{Acknowledgements}

The authors would like to thank Science and Engineering Research Board(SERB), Department of Science and Technology, Government of India for funding this work. This work is funded under the project No. IITM/SERB/SKP/154 of DST-SERB.

\end{document}